# Phonon Spectra and Phase Transitions in van der Waals Ferroics MM'P$_2$X$_6$

A. Kohutych, V. Liubachko, V. Hryts, Yu. Shiposh, M. Kundria, M. Medulych,

K. Glukhov, R. Yevych, Yu. Vysochanskii*

*Institute for Solid State Physics and Chemistry, Uzhhorod National University, Uzhhorod, Ukraine*

*corresponding author: vysochanskii@gmail.com

For CuInP$_2$S$_6$ layered ferrielectric crystal, that is most investigated among wide family of van der Waals ferroics MM'P$_2$X$_6$ (M – Cu, Ag; M' – In, Bi, Cr, …; X – S, Se), by Brillouin spectroscopy the temperature dependence of the longitudinal hypersound velocity was investigated for the acoustic phonons propagated in the plane of crystal layers. Pronounced softening of acoustic phonon branch is observed in the paraelectric phase below $T^* \approx 330$ K at cooling to temperature $T_c \approx 312$ K of the first order phase transition into ferrielectric phase. Below T$_c$ the hypersound velocity growth in correlation with spontaneous polarization rise and its temperature anomaly is similar to earlier observed temperature behavior of longitudinal ultrasound velocity for acoustic wave propagated normally to the structural layers. Determined anomalous part of the CuInP$_2$S$_6$ crystal thermal conductivity in the vicinity of ferrielectric transition also demonstrate lowering of the heat transferring phonons group velocity with cooling from $T^*$ to $T_c$. Observed acoustic softening obviously is induced by flexoelectric coupling of relaxational soft polar optic and acoustic branches that is related to inhomogeneously polarized state appearing between paraelectric and ferrielectric phases.

### Introduction

For van der Waals family of CuInP$_2$S$_6$ crystals with ferrielectric ordering below $T_c \approx 312$ K earlier the possibility of spontaneous polarization switching in samples with a thickness of several structural layers was discovered [1,2]. This discovering has opened a new direction of basic and applied research in the field of nanoelectronics based on van der Waals multiferroics [2-8]. The range of the application of $CuInP_2S_6$ crystals also is widened by their ionic conductivity – here $Cu^+$ cations are involved in both lattice spontaneous polarization normally to structural layers and charge transfer mostly along with the directions in the plane of layers [9].

The ferrielectric polarization of considered crystals is determined by opposite shifts of $Cu^+$ and $In^{3+}$ cations out of structural layers that are built by $(P_2S_6)^{4-}$ anionic structural groups. Such structural ordering below $T_c$ can be presented as freezing of $Cu^+$ cations in multiwell potential at temperature lowering. Similarly, displacive/order-disorder phase transition occurs in the $Sn_2P_2S_6$ ferroelectric crystals with three-well local potential for spontaneous polarization fluctuations [10]. In the case of $CuInP_2S_6$ crystals, the four-well (or quadruple) local potential can be involved in the description of ferrielectric ordering. The origin of such a complicated potential landscape for the copper atoms moving inside the crystal lattice elementary cells can be related to the second-order Jahn-Teller (SOJT) effect destabilizing the $d^0$ $Cu^+$ cations in positions in the middle of structural layers [11].

For $AgInP_2S_6$ and $AgInP_2Se_6$ compounds, the $Ag^+$ cations create stronger covalent bonds inside the chalcogenide octahedra preventing SOJT effect and the transition into a polar state when temperature lowered [12]. But, in $AgBiP_2S_6$ and $AgBiP_2Se_6$ compounds, the physical origin of anharmonicity can also be traced to the existence of the stereochemically active $s^2$ lone pair of $Bi^{3+}$ cations [13]. The $s$ electron shell of Bi is easily deformed by lattice vibrations, resulting in a strong anharmonicity.

In the case of the $CuBiP_2Se_6$ crystal the $Cu^+$ ions move off-center of the octahedral sites [12] and create dipole moments. However, the $Bi^{3+}$ ions within the same layer displace in the opposite direction as $Cu^+$ ones and create an almost equal but opposite dipole moment. This gives rise to the antiferroelectric arrangement and as results an intraslab dipole moment cancellation [13]. The reasons for the differences in behavior of $CuBiP_2Se_6$ (antiferroelectric) and $CuInP_2Se_6$ (ferrielectric) could be traced to the existence of the stereochemically active $s^2$ electron lone pair of $Bi^3$. The inability of $d^0$ $In^{3+}$ cations to behave similarly allows $Cu^+$ created dipole moments to induce the ferrielectric state in the $CuInP_2Se_6$ crystal.

Data of Raman scattering investigations [14] illustrate the temperature-dependent transformation of the low-frequency spectral lines intensity, which is related to $Cu^+$ cations redistribution between wells of the local potential. For further analysis of $CuInP_2S_6$ crystal anharmonic lattice dynamics, we performed DFT calculations of its electron spectra and phonon spectra in the GGA approach with considering of $s$, $p$, and $d$ valence orbitals of atoms constituting the crystal lattice [15].

Comparative analysis [15,16] of DFT calculated phonon spectra and thermal conductivity investigations [16,17] for the layered compounds $Cu(Ag)InP_2S(Se)_6$, $AgBiP_2S(Se)_6$, and $CuBiP_2Se_6$ presented an explanation of the thermal properties as ions are substituted, showing the role of disorder, electronic levels hybridization, ions size and coordination on SOJT effect. It was shown that in some compounds the thermal conductivity has very low values due to the enhancement of phonon scattering events, expressing a strong anharmonic behavior, which is justified based on interactions among optical and acoustic phonon branches as well as the presence of electron lone pairs.

Additional information on lattice dynamics and dipole ordering peculiarities can be found in consideration of acoustic phonon spectra and the temperature dependence of the sound velocity. Earlier the ultrasound properties of $CuInP_2S_6$ crystals in detail were investigated [18]. Here we present for $CuInP_2S_6$ crystal the Brillouin scattering data with the information about the temperature dependence of hypersound and analyze its thermal conductivity anomaly in the vicinity of the transition between the paraelectric and ferrielectric phases.

**Experimental results**

For grown by Bridgeman method $CuInP_2S_6$ layered crystals [19] the Brillouin scattering spectra at different temperatures were investigated with earlier described [20] equipment. By analysis of Brillouin spectra involving acoustic phonons with wave vectors oriented along $CuInP_2S_6$ crystal layers (an example is shown in Fig. 1) the temperature dependence of hypersound velocity for longitudinal (LA) and transverse (TA) waves was found. For LA phonons at 100 K the hypersound velocity $v_{LA} \approx 5330$ m s$^{-1}$ (Fig. 2a). The measured value of $v_{LA}$ agrees with the calculated velocity of LA phonons propagating in the XY plane – along crystal layers (Fig. 3b). DFT calculations of phonon spectra, with early described methodology [15], predict small elastic anisotropy of $CuInP_2S_6$ crystal in the plane of layers, but this anisotropy appears at deviation out of layers (Fig. 3a) – in normal to layers direction the longitudinal sound velocity decreased till 3500 m s$^{-1}$. Ultrasound data [18] about longitudinal sound velocity normally to structure layers of $CuInP_2S_6$ crystal (Fig. 2b) confirm high elastic anisotropy.

At heating in the ferrielectric phase, the longitudinal hypersound velocity monotonously decreases and rapidly drops above room temperature demonstrating a sharp minimum with minimal value $\approx 5050$ m s$^{-1}$ near first-order phase transition with $T_c \approx 312$ K (Fig. 2a). The minimum of $v_{LA}(T)$ dependence has pronounced asymmetry into the high-temperature side what is similar to ultrasound velocity anomaly (Fig. 2b). In the ferrielectric phase this velocity rapidly rises till ≈ 5150 m s$^{-1}$ at 300 K. Such nontrivial shape of $v_{LA}$(T) anomaly with unexpected acoustic softening in the wide temperature range (312 K – 330 K) of the paraelectric phase can't be related to the order parameter developed fluctuations or influence of the defects. The presence of some polar clusters above $T_c$ in CuInP$_2$S$_6$ crystal is supported by the second optical harmonic generation [22] indicating the crystal lattice acentricity. The $^{31}$P NMR spectral data is also illustrate [23] some residual nonequivalence of phosphorous atoms inside P$_2$S$_6$ structural groups in the range 312 K – 330 K because of their acentricity.

The thermal conductivities of MM'P$_2$S(Se)$_6$ compounds with layered crystal structure have been investigated by means of an *ac* photopyroelectric calorimetry [16,17]. In this work we analyze the thermal transport properties of the crystals in the vicinity of the ferrielectric phase transition. Thermal conductivity has been fitted by the known equation $\kappa(T) = A/T + B$, where $A/T$ is related to the three-phonon scattering processes and $B$ is the so-called lowest possible thermal conductivity in the Cahill-Pohl model [24]. Such an equation describes well the thermal conductivity in the vicinity and above the Debye temperature in case of the absence of structural phase transitions. Example of the fitting with the parameters: $\Theta_{Debye}$ = 64 K, $A$ = 109.9 W m$^{-1}$ and $B$ = 0.298 W m$^{-1}$ K$^{-1}$ is shown in Fig. 4 for AgInP$_2$Se$_6$ crystal which does not demonstrate the structural phase transitions [16].

For CuInP$_2$S$_6$ compound, with the Debye temperature $\Theta_{Debye}$ = 129 K, the fitting of earlier found [17] temperature dependence of thermal conductivity by equation $\kappa(T) = A/T + B$ in both cases, at heat transfer along and normally to structural layers (Fig. 5), deviates in the temperature vicinity of the ferrielectric phase transition, a.e. at approaching to $T_c \approx 312$ K.

The deviation $\Delta\kappa$ of the calculated dependence (red curve at Fig. 6) from the experimental data is similar for both parallelly and normally oriented thermal flows and has the shape of the asymmetric minimum that is positioned at $T_c$ (Fig. 7). These deviations can be compared with the temperature dependencies of the optical birefringence increment $\Delta n(T)$ [25] and with the square of a spontaneous polarization $P_s^2$ of the ferrielectric phase [26]. Temperature anomaly of the longitudinal ultrasound velocity in normal to crystal layers direction [18] and determined by Brillouin scattering spectroscopy temperature anomaly of the longitudinal hypersound velocity, propagated in the plane of the crystal layers, also are compared with thermal conductivity deviation related to the ferrielectric phase transition.

**Discussion of results**

Anomalous lowering of thermal conductivity in $CuInP_2S_6$ crystal together with the decreasing of the ultrasound and hypersound velocities can be considered as the evidence of heat transferring acoustic phonons group velocity decrease in the paraelectric phase at cooling to $T_c$. Observed temperature dependence of longitudinal ultrasound and hypersound below $T_c$ correlates with the square of spontaneous polarization and agrees with Landau – Khalatnikov theory for acoustic anomalies at ferroelectric ordering [27]. Lowering of sound velocity in the wide temperature range of paraelectric phase at cooling to $T_c$ can be related to the longitudinal acoustic phonon branch suppression by softened optical relaxational modes of spontaneous polarization fluctuations in the case of order - disorder type of the phase transition occurring in $CuInP_2S_6$ crystal. Such overdamped (relaxational) optical mode is clearly observed by dielectric microwave investigations [18]. The linear interaction of optical and acoustic phonon branches is determined by flexoelectric coupling and can be described by an accounting of the Lifshitz like invariants in the thermodynamic potential for the monoclinic symmetry of $CuInP_2S_6$ crystal, similarly, as was done at the analysis of sound velocity anomalies near the Lifshitz point (LP) on temperature – composition diagram of $Sn_2P_2(Se_xS_{1-x})_6$ ferroelectrics [21].

In the case of $Sn_2P_2(Se_xS_{1-x})_6$ ferroelectrics the development of space inhomogeneous fluctuations of spontaneous polarization at cooling to the edge of the paraelectric phase stability is related to the incommensurate phase appearance between the paraelectric and ferroelectric phases for compositions $x > x_{LP}$ [21]. For layered ferrielectrics $CuInP_2S_6$ the acoustic and thermal conductivity investigations show that some specific temperature region between $T_c \approx 312$ K and $T^* \approx 330$ K is inclined between the paraelectric and ferrielectric phases. Here the cooperative behavior may be envisioned in terms of dipole - dipole correlations developing with freezing of random $In^{3+}$ $d^{10}$ cations motions and with the ordering of $Cu^+$ $d^{10}$ cations random positions. If the dipoles are part of and orthogonal to the $CuInP_2S_6$ crystal plane, the appearance of long-range order would involve antiparallel displacements which minimize the electrostatic and elastic energy costs of ordering.

According to structural investigations [26] in $CuInP_2S_6$ crystal lattice above $T_c$, the cooper cations are disordered around positions at top and bottom of crystal layers with occupation probabilities symmetrical relatively second order symmetry axis $C_2$, which is placed at the middle of structural layer and oriented along Y (*b*) axis of monoclinic (*C2/c*) elementary cell. At this, the indium cations are also distributed among positions that are a little shifted above or below of $C_2$ axis. In the case of antiferroelectric interaction between nearest-neighbor dipoles (in the plane of the given layer) created by indium cations shifts, and at the ferroelectric coupling of the nearest dipoles related to cooper cations positions in neighbor structural layers, some frustrations in the dipole space ordering can appear. Such possibility is supported by dipole glassy type disordering at low temperatures inside the ferrielectric phase of $CuInP_2S_6$ crystals [28]. For the $CuInP_2Se_6$ compound the dipolar frustration was supposed at the interpretation of successive phase transitions according to the observed temperature dependence of dielectric susceptibility [29]. The frustration was appeared by piezoelectrically active domain walls between competitive ferrielectric and antiferroelectric states [30,31]. Also, the dipole glassy state in the mixed crystal $CuInP_2(Se_xS_{1-x})_6$ [28] obviously is related to frustration effects at dipole ordering.

The DFT calculations [15] illustrate the hybridization of valence orbitals for $Cu^+$ and $In^{3+}$ cations in $d^0$ nominal state. The local SOJT effect strongly displaces the $Cu^+$ cations from the center of the surrounding sulfur octahedron (such effect is a little weakened in selenide compound). For the indium cations in $d^0$ electronic configuration, the acentricity inside chalcogen's octahedrons is relatively small. But due to the high charge of $In^{3+}$ cations appeared dipoles are big enough for partial compensation of electric dipoles related to the $Cu^+$ position. It can be supposed that at disordering of $Cu^+$ cations above $T_c$, in some temperature interval – till $T_c$ + 30 K, the antiferroelectric type ordering for $In^{3+}$ sublattice still exists. Such possibility can be described in the mean-field approximation suggesting that an order parameter of the ferrielectric transition in considered crystals interacts with another order parameter, which is related to the structural transition – obviously with some structure modulation. Such interaction is enough strong for $CuInP_2S_6$, and here, at temperature decreasing, the second-order and first-order transitions can occur. For $CuInP_2Se_6$ crystals named interaction evidently is weak that can determine at cooling in a series of the second-order phase transitions [29].

In the pseudospin quantum anharmonic oscillators model with two sublattices (one of them with Ising like two – well local potential, another with Blume–Emery–Griffiths model like three – well local potential at given single-ion anisotropy) and with the account of competition between nearest and next – nearest neighbors interactions the ground state and the temperature – pressure – composition phase diagram of layered $CuInP_2S(Se)_6$ ferri-(antiferro-)electrics can be described, similarly to elucidation of dipole ordering in case of $Sn(Pb)_2P_2S(Se)_6$ ferro-(antiferro-)electrics with three-dimensional crystal lattice [32,33].

**Conclusions**

The Brillouin spectroscopy and thermal conductivity investigations confirm earlier received ultrasound data [18] about unexpected elastic softening of the $CuInP_2S_6$ layered crystals at the edge of stability of the paraelectric phase – at cooling to the first order transition into the ferrielectric phase. The observed effect at the high-temperature side of the order – disorder type phase transition can be induced by flexoelectric coupling of relaxational soft polar optical and acoustic branches. Such coupling can induce an inhomogeneously polarized state in the temperature region between the paraelectric and ferrielectric phases. This peculiarity is very important for $MM'P_2X_6$ family ferroic materials, which is already widely integrated into different functional heterostructures, and requires further theoretical and experimental study.

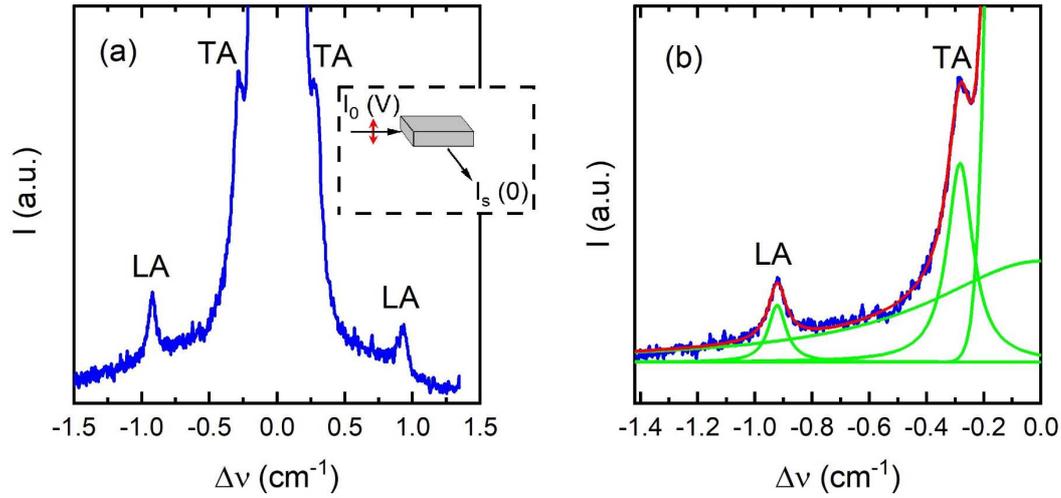

Figure 1. Brillouin spectrum for $CuInP_2S_6$ crystal at 295 K with light scattering geometry involving acoustic phonons propagating in the plane of crystal layers (a). The spectral lines of longitudinal (LA) and transverse (TA) acoustic phonons fit by Lorentz's peak function (b). Lorentz contour of relaxation dynamics and elastic scattering Gaussian peak are also shown.

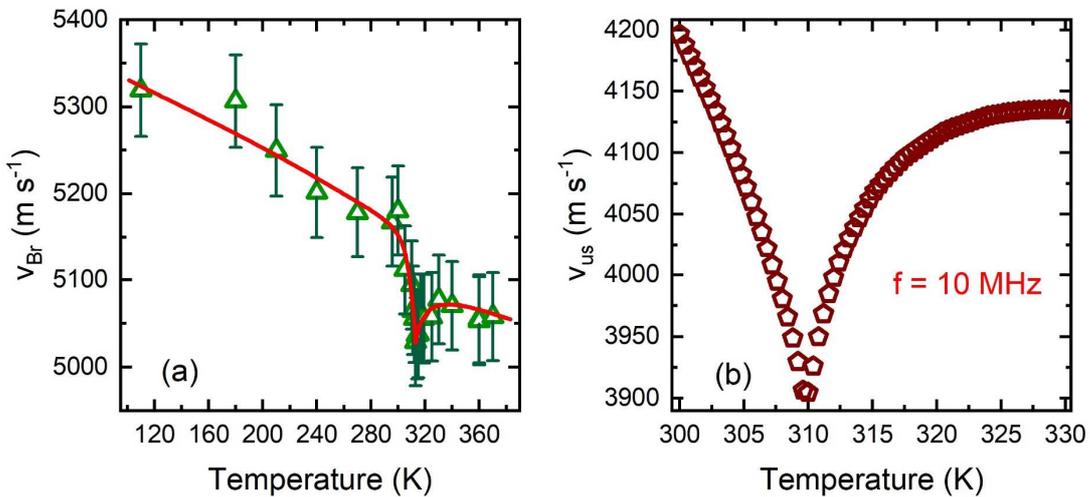

Figure 2. Temperature dependence of longitudinal hypersound velocity with the acoustic phonons propagated in the plane of the crystal layers (a), and longitudinal ultrasound propagated normally to the crystal layers [18] (b) for $CuInP_2S_6$ crystal.

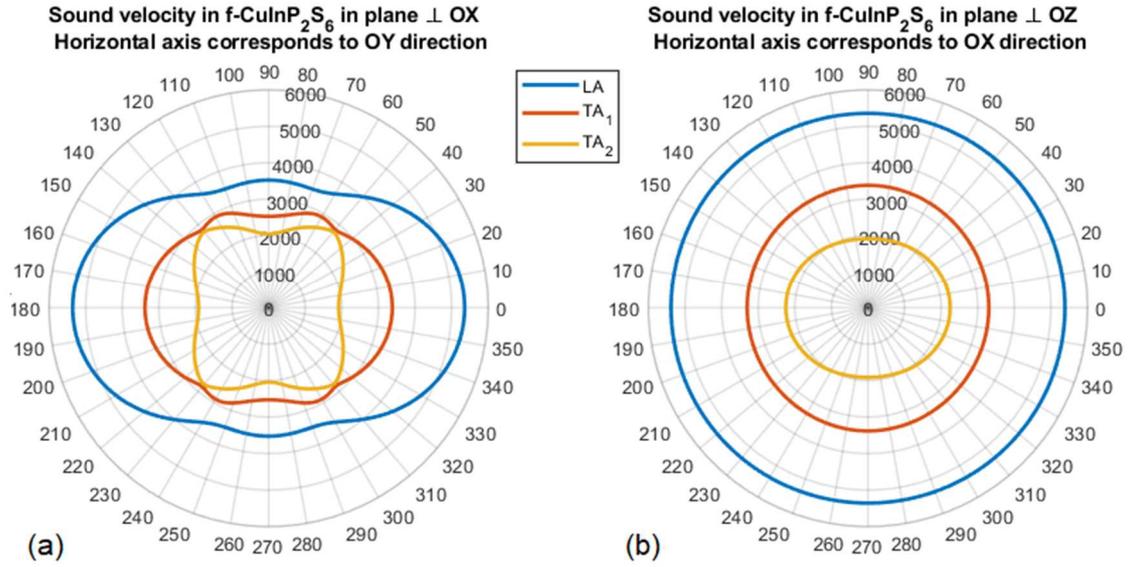

Figure 3. DFT calculated for CuInP$_2$S$_6$ ferrielectric phase sound velocity angle dependencies in Cartesian plane YZ oriented normally to crystal layers (a) and in coincided with crystal layers plane XY (b).

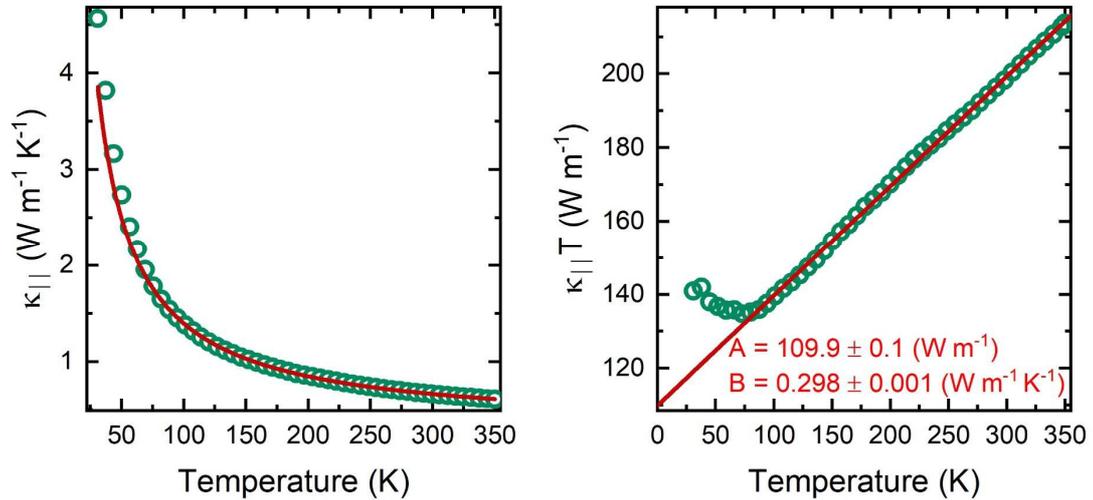

Figure 4. Temperature dependence of thermal conductivity of AgInP$_2$Se$_6$ crystal and such dependence in coordinates $T\kappa$ versus $T$ (left). The solid line is obtained by fitting with equation $\kappa(T) = A/T + B$.

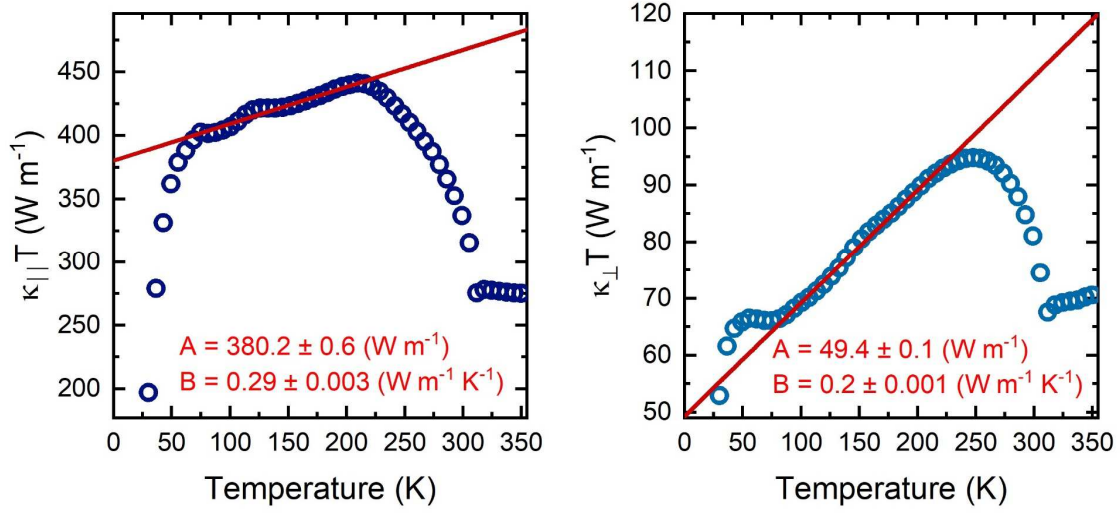

Figure 5. Temperature dependence of CuInP$_2$S$_6$ crystal thermal conductivity κ measured along (left) and normally (right) to the structural layers in coordinates $T$κ versus $T$. The red line was calculated using eq. κ ($T$) = $A/T + B$.

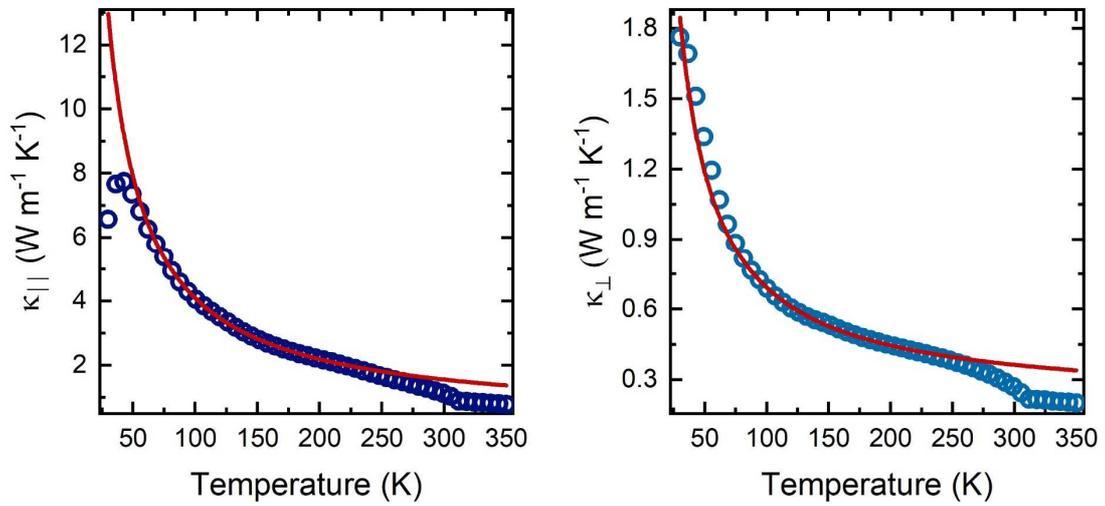

Figure 6. Temperature dependence of thermal conductivity of CuInP$_2$S$_6$ crystal measured along (left) and normally (right) to the layers. The red curve was calculated using eq. κ = $A/T + B$.

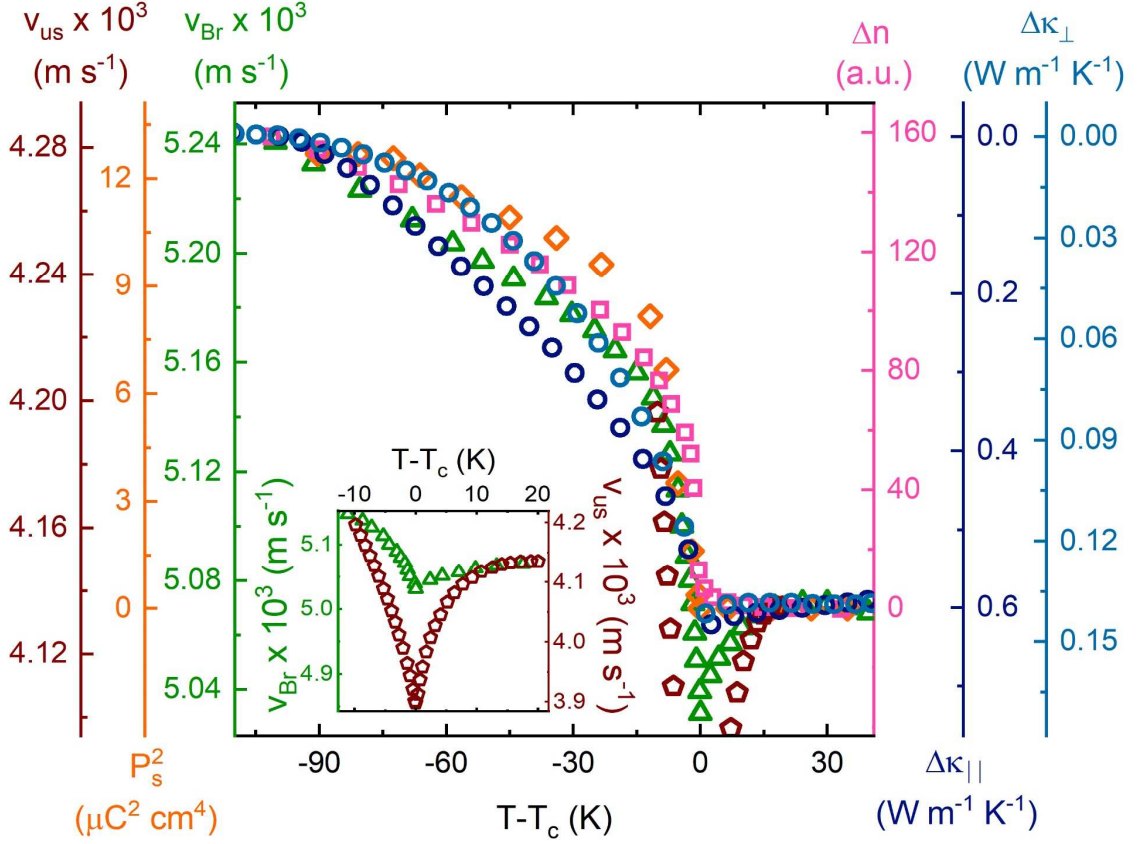

Figure 7. Comparison of temperature anomalies of thermal conductivity deviations $\Delta\kappa$ (circles) along and normally to the structural layer of $CuInP_2S_6$ crystal, with temperature behavior of the optical birefringence increment $\Delta n$ (squares) [25], the square of spontaneous polarization $P_s^2$ (rhombuses) [26], longitudinal ultrasound velocity $v_{us}$ normally to the layers (pentangles) [18] and hypersound velocity $v_{Br}$ along the crystal layers (according to our Brillouin scattering data) (triangles) in the vicinity of the ferrielectric phase transition.